\documentclass{ws-ijbc}
\usepackage{ws-rotating}     
\usepackage{subfigure}
\begin{document}

\catchline{}{}{}{}{} 

\markboth{Gambuzza, Fortuna, Frasca, Gale}{Experimental evidence of chaos from memristors}

\title{Experimental evidence of chaos from memristors}

\author{LUCIA VALENTINA GAMBUZZA, LUIGI FORTUNA, MATTIA FRASCA}
\address{DIEEI, University of Catania, Viale A. Doria, 6\\
Catania, 95125, Italy\\
mfrasca@diees.unict.it}

\author{ELLA GALE}
\address{Bristol Robotics Laboratory, Coldharbour Lane, Bristol, BS16 1QY Email: ella.gale@brl.ac.uk\\
Applied Mathematics \& Sciences, Khalifa University of Science, Technology and Research,\\ Abu Dhabi, United Arab Emirates, PO Box 127788
}

\maketitle

\begin{history}
\received{(to be inserted by publisher)}
\end{history}

\begin{abstract}
Until now, most memristor-based chaotic circuits proposed in the literature are based on mathematical models which assume ideal characteristics such as piece-wise linear or cubic non-linearities. The idea, illustrated here and originating from the experimental approach for device characterization, is to realize a chaotic system exploiting the non-linearity of only one memristor with a very simple experimental set-up using feedback. In this way a simple circuit is obtained and chaos is experimentally observed and is confirmed by the calculation of the largest Lyapunov exponent. Numerical results using the Strukov model support the existence of robust chaos in our circuit. This is the first experimental demonstration of chaos in a real memristor circuit and suggests that memristors are well placed for hardware encryption.
\end{abstract}


\section{Introduction}

\label{sec:intro}

The memristor is the first (and possibly the only) non-linear fundamental circuit element, and, as such, has a lot to offer to those interested in non-linear dynamics. By virtue of being a fundamental element, we can be reasonably certain that it's the simplest non-linear circuit element. It is a physical device, announced to the world in 2008~\cite{strukov08}, although its existence was predicted many years earlier in 1971~\cite{Chua1971}. The non-linear behaviour arises because the memristor relates the total history of the charge (the time integral of current), that has passed through the device, with the total history of the magnetic flux (the time integral of voltage)~\cite{Chua1971}.

Since the first physical realization of a memristor \cite{strukov08} a huge amount of effort has been devoted to the study of materials and manufacturing techniques for memristor devices, as well as theoretical models, (as is covered in detail in a recent review~\cite{RevMemReRAM}). In terms of materials, the first and archetypal memristor is a device made of titanium dioxide sandwiched between platinum electrodes. Studies on TiO$_2$ memristor~\cite{strukov08} have shown that the main cause of the memristive effect is the motion of oxygen vacancies from the high resistance layer, the TiO$_2$ layer, to the sub-oxide (TiO$_{2-x}$), or low resistance layer, under the application of an external bias: it is this interconversion that changes the resistance of the device. The realization of the active layer requires expensive techniques, such nano-imprint lithography and atomic layer deposition, followed by subsequent annealing at high temperature. Most memristors are fabricated by atomic layer deposition as perfect crystals and annealing is usually required to introduce vacancies, as these are errors in a perfect crystal. The fabrication techniques significantly influence the device performance and significantly impact the cost-effectiveness of the device. For this reason, new methods, typical of the printing industry, screen printing or ink-jet printing, have been also investigated for the realization of memristive devices \cite{duraisamy,memZno}. Despite the numerous realizations of memristors, a breakthrough in ease of manufacturing was the fabrication of a device by spin-coating a titanium isopropoxide solution on a flexible plastic substrate \cite{hackett}. Unlike devices created by atomic deposition, sol-gel and ink-jet printed devices already have vacancy errors, so they require no annealing step for the formation of the active layer. 
In this paper, we make use of flexible TiO$_2$ memristors \cite{ella2011}.



Memristors belong to the class of resistive switching devices, which in recent years, attracted a lot of interest for memory, logic and neuromorphic applications. In particular, the memory effect is due to the possibility to switch the status of the device, through SET and RESET processes, between two states, the high resistance state (HRS), and the low resistance state (LRS). In bipolar RRAM, the SET and RESET processes occur through the formation/dissolution of conductive filaments \cite{ielmini2011}. This mechanism is controlled by a compliance current. Depending on the compliance current, during the switching of a metal-insulator-metal structure between the HRS and the LRS, random telegraph noise (RTN) may be observed. RTN can be reduced with proper SET and RESET processes, but, on the contrary, fluctuations of current at different resistance states can be exploited for non-destructive studies of switching phenomena at microscopic scale \cite{ielmini10}. RTN is also used as the principle for true random number generators based on a contact resistive RAM \cite{Huang}.

As resistive switching devices, the memristors are also of interest for implementation of Boolean logic gates \cite{borghetti2010,Rose2012,Ella2013,Kvatinsky2014,Vourkas2014}, for applications as nonvolatile memories \cite{nonvolatileMem2011}, and as synapses in neuromorphic circuits \cite{Jo2010}.
Due to its non-linearity, the use of the memristor as nonlinear component in chaotic circuits \cite{muthuswamy10} has been envisaged. To this aim the memristor has been used to substitute for Chua's diode in versions of Chua's circuits \cite{itoh08,muthuswamy10,253} and simulated chaos has been observed \cite{82,61,70,232}. Most of the memristor-based oscillators used in these chaotic circuit simulations assume ideal characteristics for the memristor, \emph{e.g.} cubic or piece-wise-linear (PWL) nonlinearities \cite{itoh08}. Since the Strukov memristor is a passive non-symmetrical element having a non-linearity different from the one most frequently investigated, its use for chaotic circuits design is not trivial. Specifically, the non-linearity assumed in literature is usually a continuously varying function or a piece-wise linear-continuously varying function for filamentary type devices.

We would like to demonstrate experimentally that the memristor can be used as the non-linear element in a chaotic circuit. The first step is to demonstrate that simulated circuits including a realistic model of memristor exhibit chaos, this has been done in ~\cite{buscarino12,Buscarino2013} where recently a gallery of chaotic circuits using the HP model has been proposed: these papers utilised a particular configuration of two memristors connected in anti-parallel, which allowed the authors obtain a symmetrical characteristic suitable for chaos generation. This sub-circuit has been experimentally realised in ~\cite{spike} where circuits made of 2 or 3 memristors were found to give rise to complex dynamics. What has not been done is to conclusively demonstrate chaotic dynamics in an experimental circuit containing a memristor and to do so with only one non-linearity in the circuit (i.e. only one memristor). In our experiments we do not apply specific SET and RESET processes, but we drive the memristor by a signal which is function of the actual status of the device, exploiting its switching properties to obtain deterministic chaos.

The aim of this work is to provide an experimental demonstration of chaos with memristive devices. For this purpose, a simple experimental set-up is proposed in Sec.~\ref{sec:modelA}, where chaos can be uniquely attributed to the presence of the memristor, as described in Sec.~\ref{sec:resultsA}. We then perform a numerical simulation of the set-up as described in Sec.~\ref{sec:modelB} using the simplest model available and explore the parameter space to demonstrate that the chaos is a robust phenomenon observed over a large parameter space in Sec.~\ref{sec:resultsB}. In Sec.~\ref{sec:conclusion} the conclusions of the work are drawn.

\section{Methodology}
\label{sec:model}

\subsection{Experimental}
\label{sec:modelA}


The idea underlying the experimental set-up originates from the way in which memristors are usually characterized. One of the main fingerprints of memristors is the pinched hysteresis loop in the $v-i$ plane measured under periodic excitation \cite{Chua2011,Biolek2013}. A commonly used technique to obtain the $v-i$ characteristics is to apply a sinusoidal voltage input to the memristor (alternatively, the time response is studied through the application of pulse functions) and to measure the current through the device. To do this, a (Keithley 2602) programmable sourcemeter is commonly used because it allows the tuning of the input parameter and the recording of the current. The core of this work is to dynamically change the parameters of the input waveforms as a function of the actual state of the memristor, to our knowledge, this is the first experiment using such a feedback loop.

The memristor used in this work has a sandwich structure of Al-TiO$_2$-Al: the aluminium electrodes were sputter-coated onto PET plastic substrate via a mask as in \cite{hackett,ella2014}, the sol-gel layer \cite{ella2014} was deposited via spin-coating (spun at 33r/s  for 60s) and then left under vacuum to hydrolyse for an hour as in \cite{ella2011}. The electrodes are 4mm wide, crossed at 90$^{\circ}$, giving an active area of 16mm$^2$, the titanium gel layer is 40nm thick. In contrast to other more expensive techniques for the fabrication of the active layer, the spin-coated memristor was created with a simple procedure, and without the need for a forming step.

The experimental methodology consists of two separate steps. In the first, we test the memristive behavior by exciting it with a periodic input and observing the behavior in the $v-i$ plane. In the second step, we drive the memristor with pulses which are related to the actual state of the memristor through a law discussed in details below. The first step corresponds to operating the memristor in open loop conditions, while the second in feedback mode. In feedback conditions the whole system operates without external inputs, so the actual status of the system (which includes the memristor status and the voltage applied to it) is the result of an autonomous evolution.

The memristor characterisation analysis (first step of our methodology) was performed by setting the sourcemeter to run linear voltage sweeps between a range of $\pm 1V$, with a voltage step size of $0.02V$ and a settling time of $0.01s$ (settling time is the delay period after which the measurement is made). The signal may be viewed as the DC equivalent of an AC waveform with a frequency of $0.5Hz$. The measurements have been repeated several times to test the device performance over iterated cycles.

\begin{figure}
\centering {{\includegraphics[width=8.5cm]{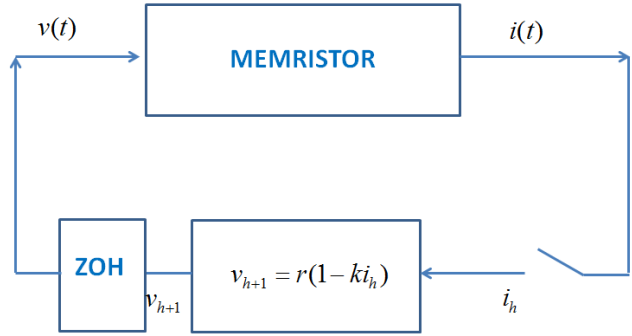}}}
\caption{\label{fig:schema2} The memristor based circuit.}
\end{figure}

In the second step, we investigate the effect of establishing a relationship between the measured current and the next sample of the applied voltage signal, that is, we drive the memristor on the basis of the current flowing through it. This is particularly simple to realize since it only requires a memristor and a programmable sourcemeter.
The scheme adopted to test for non-linear dynamics is illustrated in Fig.~\ref{fig:schema2}, where $v(t)$ indicates the voltage applied to the memristor, i.e. a digital-to-analog converted signal. The memristor current, indicated as $i(t)$, is sampled as a sequence of measurements $i_h$. Each sample $i_h$ is used to generate the next voltage sample $v_{h+1}$ through the relation:

\begin{equation}
\label{eq:vh1}
v_{h+1}=r(1-k i_h)
\end{equation}

\noindent where $k$ and $r$ are constant tunable parameters. The sequence of samples $v_{h}$ is converted into the analog signal $v(t)$ through the zero-order hold (ZOH), so that:

\begin{equation}
\label{eq:vt}
\begin{array}{ll}
v(t)=v_h, & t_h \leq t < t_{h+1}
\end{array}
\end{equation}

\noindent with $t_h=h\Delta T$, where $\Delta T$ is the sampling time. The sampling of $i(t)$ occurs immediately before the sweep of the voltage from $v_h$ to $v_{h+1}$, that is, $i_h=i(t)|_{t=t_{h+1}^-}$. We note that Eq. (\ref{eq:vh1}) is linear, so that any non-linearity in the system comes from the memristor.

The sampler, the processor implementing Eq. (\ref{eq:vh1}) and the ZOH are all implemented in the sourcemeter. During operations in feedback conditions, in the sourcemeter Eq. (\ref{eq:vh1}) is iterated, so that the voltage levels are automatically generated for a given time window during which a number of samples of the waveforms is acquired. Specifically, we used a Keithley 2602 programmable sourcemeter to apply a step-wise sweep voltage (both the step amplitude and the duration can be programmed). We used steps with fixed duration so that the applied waveform is a continuous-time signal generated by the conversion of a discrete-time signal through a zero-order hold. The measurement is made at each step after a specified delay period (the sampling time $\Delta T$). Lyapunov exponents of the data were calculated using the TISEAN package \cite{tisean}, which is based on Rosenstein's algorithm \cite{Rosenstein}.

\subsection{Simulational}
\label{sec:modelB}

\begin{figure}
\centering {{\includegraphics[width=8.5cm]{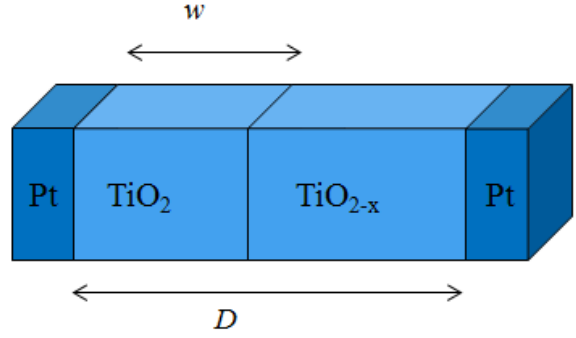}}}
\caption{\label{fig:HPmem} Schematic representation of a TiO$_2$ memristor.}
\end{figure}

Many models of memristors, ranging from very simple ones to ones incorporating a very detailed level of description of the phenomena occurring in the device, have been proposed in literature. The first electronic engineering model of a memristor was Chua's original model in 1971 \cite{Chua1971}, this model offered a way to relate the circuit measurables to the state of the system, but was too abstract to model specific systems. The first materials science model of a memristor was in the original Strukov paper \cite{strukov08}, this included a very simple approach based on modelling the memristor as a space-conserving variable resistor, where the boundary, $w$, between TiO$_2$ and TiO$_{2-x}$ (Fig.~\ref{fig:HPmem}) is used as a state variable which holds the memory of the device (this fits the chemistry of the system, as when the voltage is removed the oxygen vacancies do not dissipate like electrons or holes do in semiconductors, but remain `remembering' a state). The Strukov model assumes a linear dopant drift under a finite uniform field, which is widely thought to be unlikely in thin-film~\cite{86}. Several papers have attempted to improve on this, by introducing a window function to prevent the system from going outside of its bounds 0 and D, the thickness of the device, and slowing the boundary down near the edges \cite{2,46,306}. Nonetheless, the original model allows ease of modelling and simplicity to the interpretation of results as long as the simulation code takes care to avoid $w$ taking un-physical values, and has been widely adopted for simulations \cite{345,346}. More realistic models have also been proposed, an example is \cite{220} which gives good agreement with the experimental data, but requires 8 fitting parameters and contains difficult to simulate terms.

\begin{figure}
\centering {{\includegraphics[width=8.5cm]{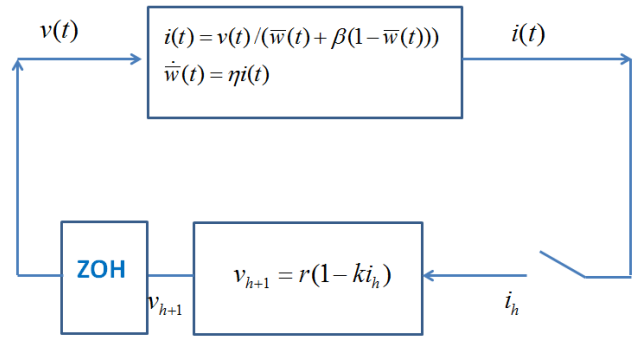}}}
\caption{\label{fig:schema1} Scheme used for numerical analysis of the memristor based circuit.}
\end{figure}

We show here that the presence of chaos is demonstrated even with a very simple model capturing only the main characteristics of the memristive behavior. Specifically, numerical results are obtained by substituting the memristor in the scheme of Fig.~\ref{fig:schema2} with its modelling equations as proposed by Strukov et al. \cite{strukov08}: the resulting simulational scheme is shown in Fig.~\ref{fig:schema1}. The relationship between current and voltage in the memristor is governed by:

\begin{equation} \label{eqn:hp model}
i(t)=v(t)/ \left(  R_{ON}{w(t)\over{D}}+{R_{OFF}\left( {1-{w(t)\over{D}}} \right)} \right) \: .
\end{equation}

The variable $w(t)$ (the width of the doped region) represents an internal memory variable limited to values between zero and $D$, the thickness of the device, and is characterized by the following dynamics:

\begin{equation} \label{eq:w2}
\dot{w}(t)={\eta \frac{\mu_v R_{ON}}{D}} i(t)
\end{equation}

\noindent where $\eta$ characterizes the polarity of the memristor ($\eta=1$ if the doped region is expanding, $\eta=-1$ otherwise), $\mu_v$ is the oxygen vacancy ion mobility, and $R_{ON}$ ($R_{OFF}$) is the resistance of the device in its lowest (highest) resistance state.
Following \cite{strukov08} Eqs.~(\ref{eqn:hp model}) and (\ref{eq:w2}) may be rescaled to obtain:

\begin{equation} \label{eqn:hp model2}
i(t)=v(t)/ \left(  \bar{w}(t)+{\beta\left({1-\bar{w}(t)}\right)} \right )
\end{equation}

\noindent and

\begin{equation} \label{eq:w2_2}
\dot{\bar{w}}(t)={\eta} i(t)
\end{equation}

\noindent where $\beta=R_{OFF}/R_{ON}$ and time is now expressed in the units of $t_0=D^2/\mu_v v_0$, with $v_0=1V$. The numerical simulations have been based on Eqs. (\ref{eqn:hp model2}) and (\ref{eq:w2_2}), which are dimensionless and contain only one parameter ($\beta$).

\section{Results}
\label{sec:results}

\subsection{Experimental results}
\label{sec:resultsA}
\begin{figure}
\centering {{\includegraphics[width=8.5cm]{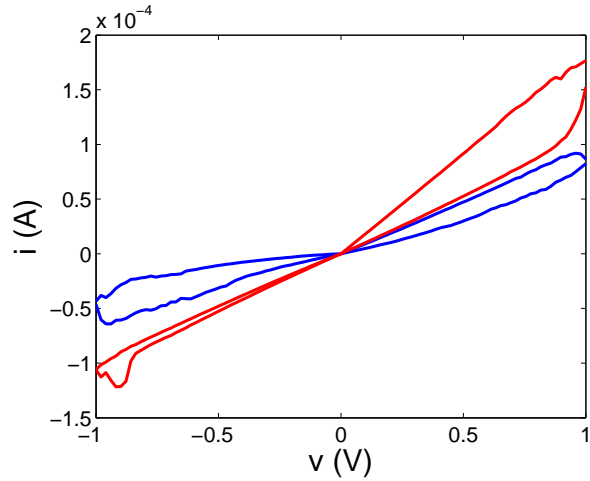}}}
\caption{\label{fig:v-iMv47} Two $v-i$ curves for one of the memristors (memristor 1) used in the experiments (similar results were seen for memristor 2 and are not shown). The input signal is a linear voltage sweep between a range of $\pm 1V$, with a voltage step size of $0.02V$ and a settling time of $0.01s$, that is the DC equivalent of an AC waveform with a frequency of $0.5Hz$.}
\end{figure}

The experimental results are reported for two devices, indicated as memristor 1 and memristor 2. The two memristors were fabricated with the same procedure, however parametric tolerances due to the fabrication process do appear. We first describe the $v-i$ characterization under periodic excitation of the memristors analyzed in this work. The results, which refer to memristor 1, are reported in Fig.~\ref{fig:v-iMv47}, showing that the resistance change is not fully reversed over the range of the hysteresis loop. However, in view of the application as chaos generator, this does not represent a severe issue, since the non-linearity is not required to be time-invariant. Similar behavior has been obtained for memristor 2.

\begin{figure}
\centering
\subfigure[]{\includegraphics[width=8cm]{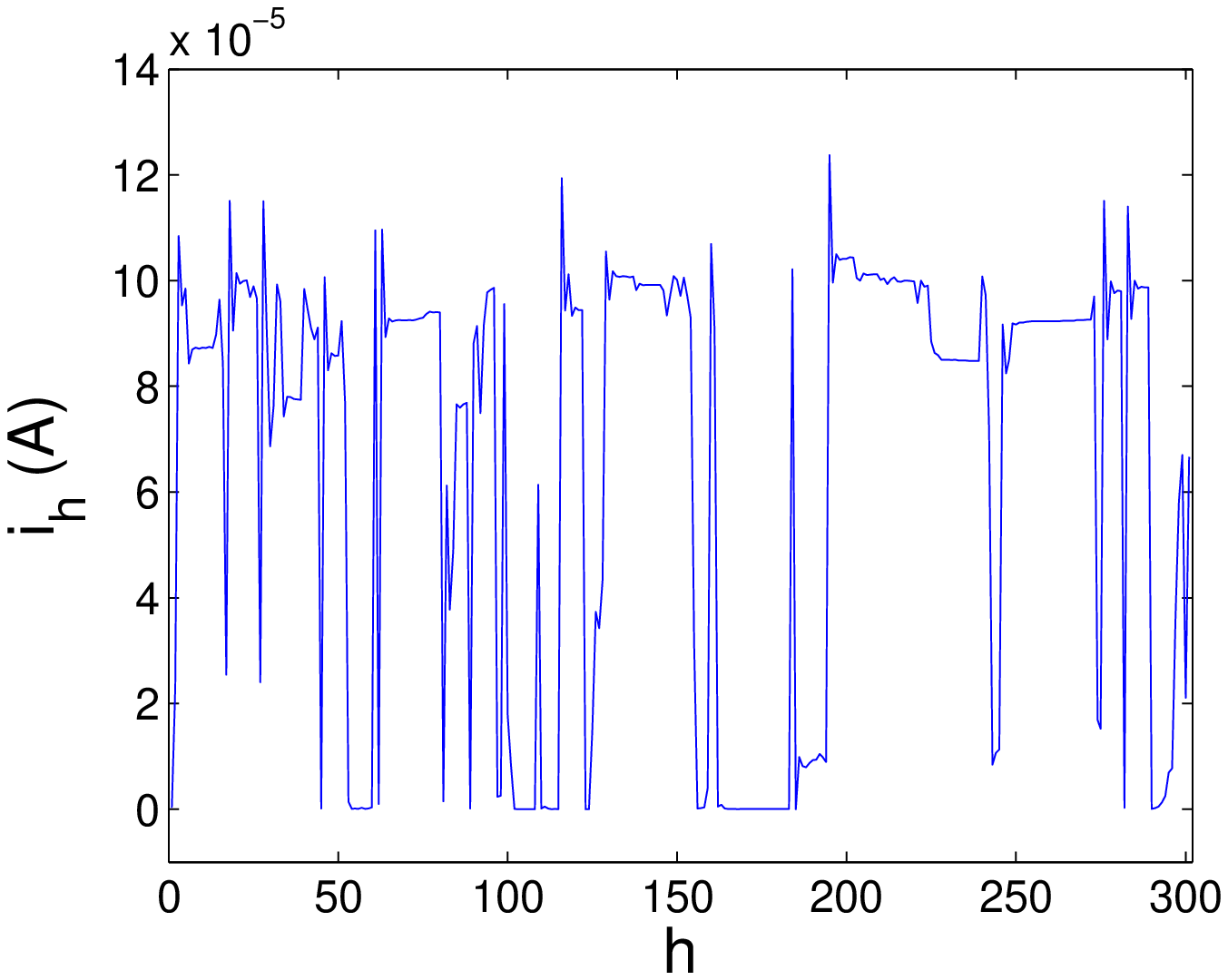}}
\subfigure[]{\includegraphics[width=8cm]{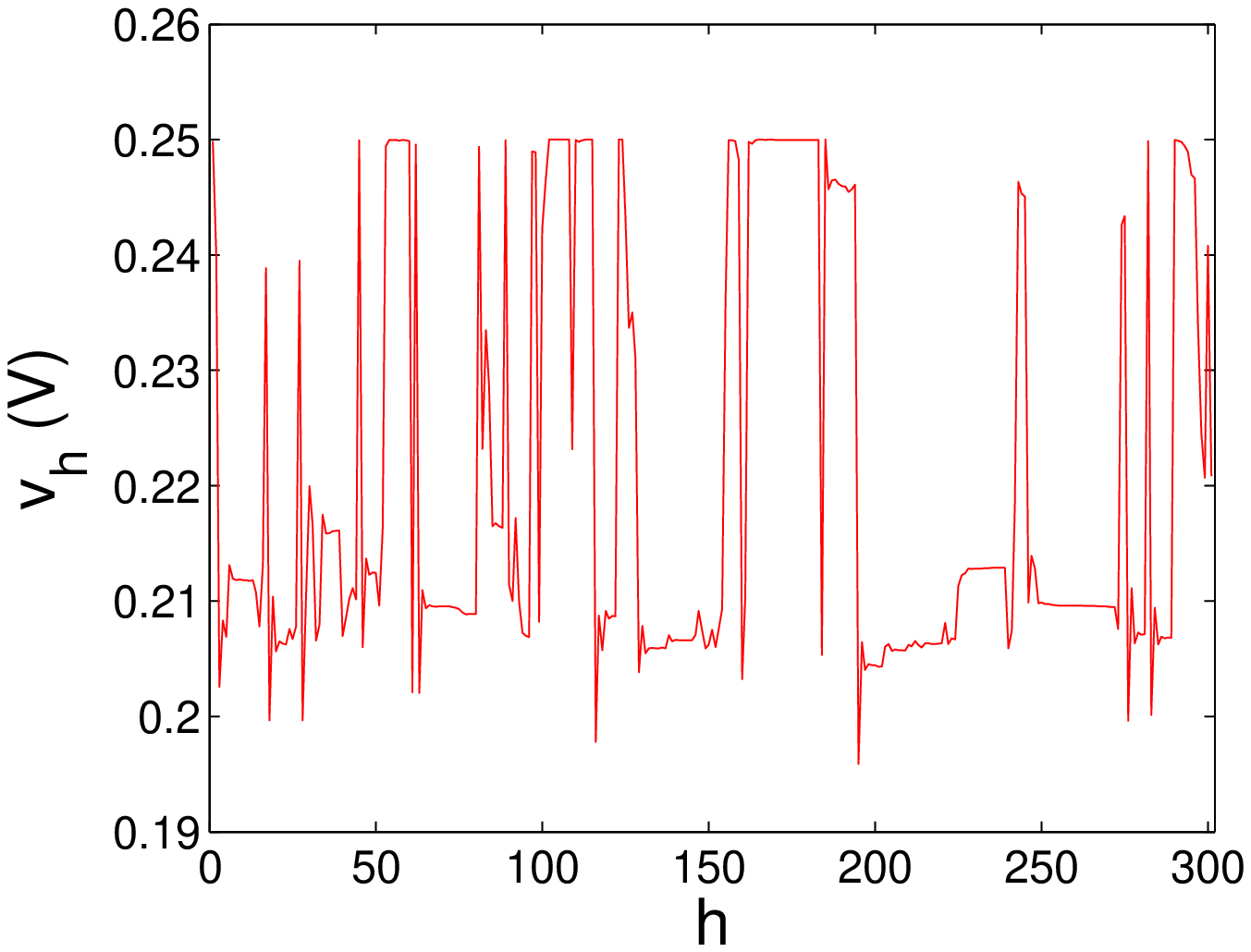}}\\
\caption{\label{fig:timeevolutioniandv2} Experimental results: time evolution of current and voltage in the memristor 1 based circuit for $r=0.25$ and $k=1750$.}
\end{figure}

\begin{figure}
\centering
\subfigure[]{\includegraphics[width=8cm]{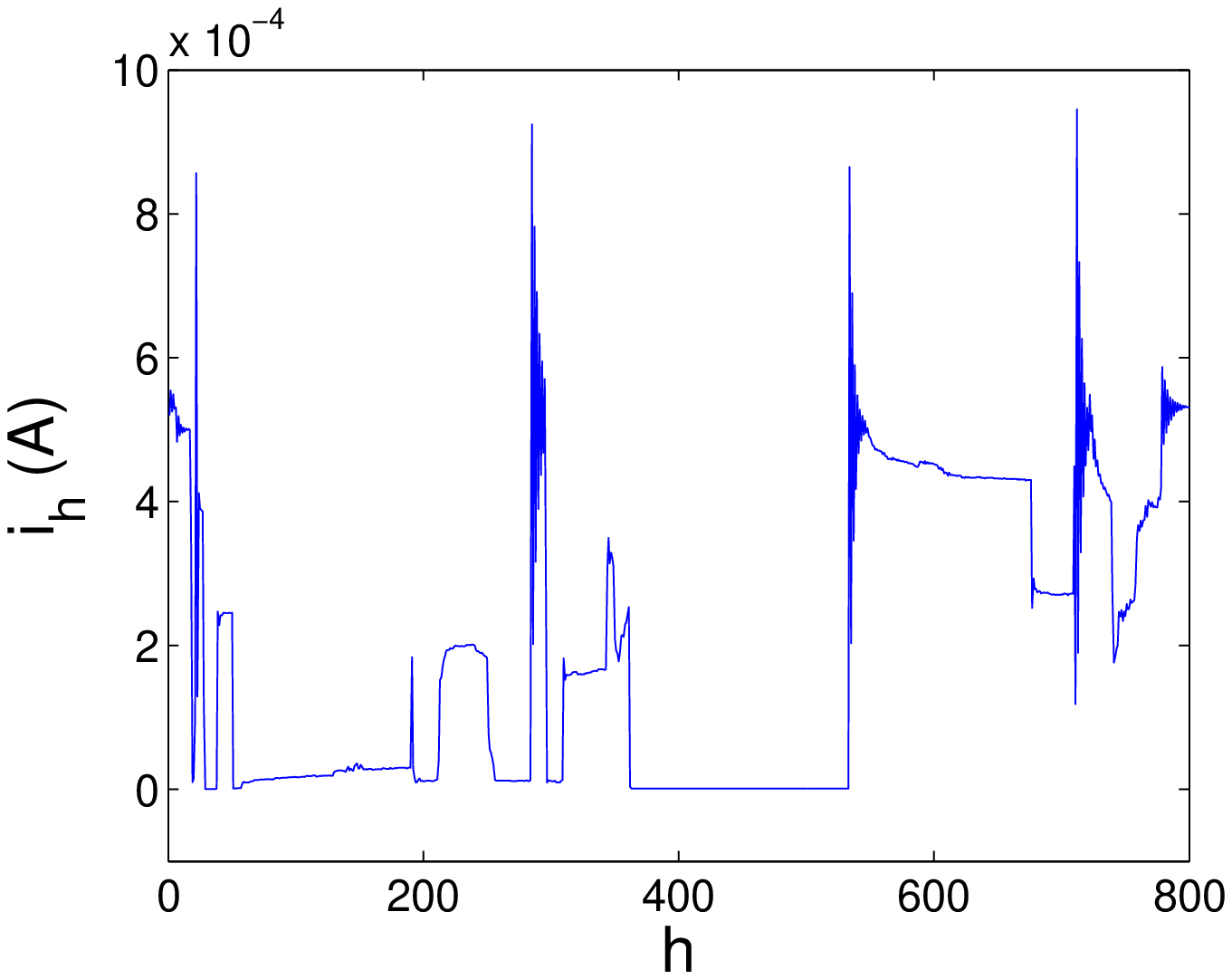}}
\subfigure[]{\includegraphics[width=8cm]{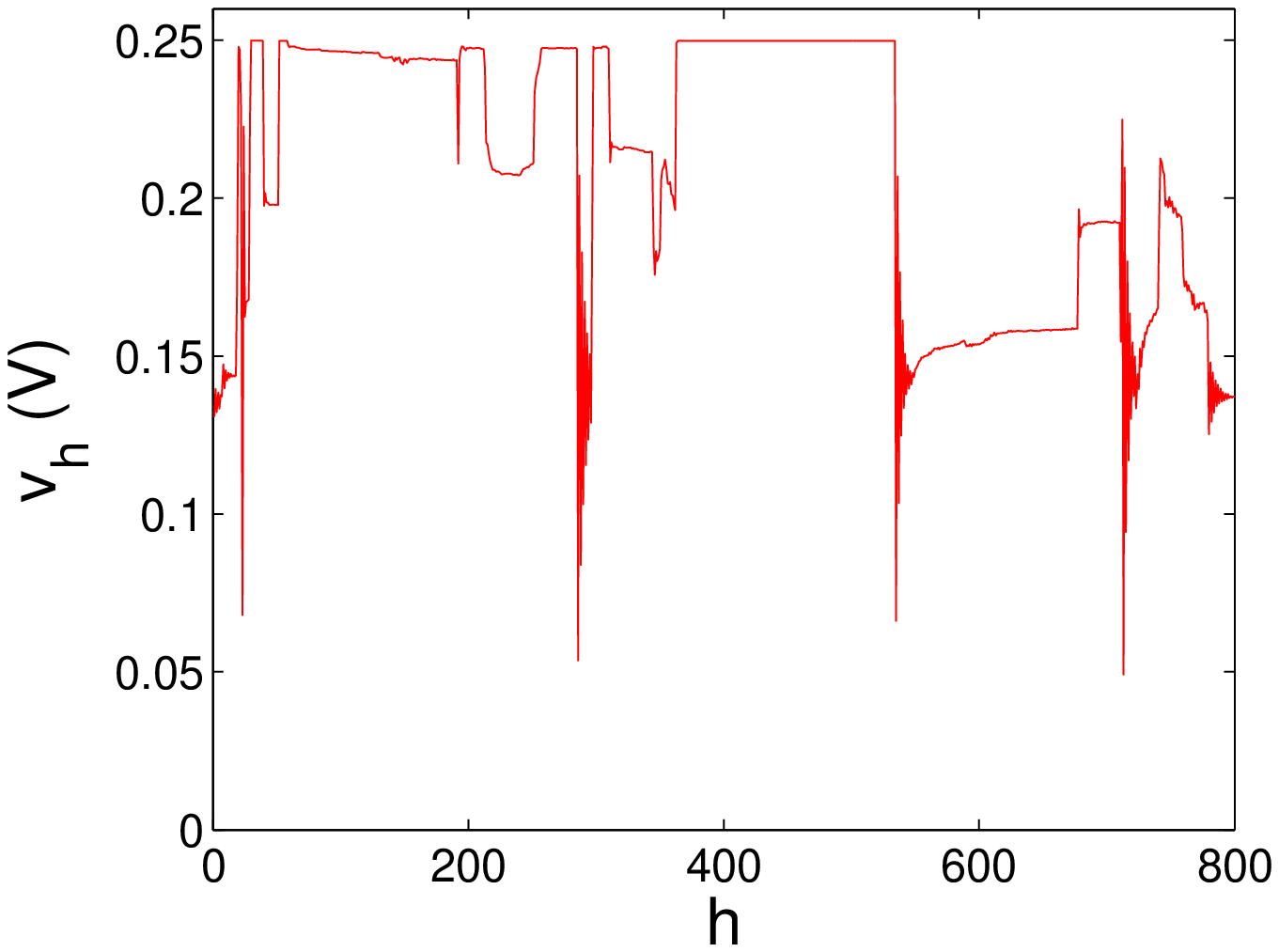}}\\
\caption{\label{fig:timeevolutioniandv} Experimental results: time evolution of current and voltage in the memristor 2 based circuit for $r=0.25$ and $k=850$.}
\end{figure}

The spin-coated TiO$_2$ memristor was then controlled by a feedback signal generated by using the scheme of Fig.~\ref{fig:schema2}. In both cases chaos has been observed, but the two devices required a different parameter tuning of the control law (\ref{eq:vh1}). For memristor 1 we have chosen $r=0.25$ and $k=1750$ and obtained the waveforms shown in Fig.~\ref{fig:timeevolutioniandv2}. For memristor 2 we have chosen $k=850$ and $r=0.25$; the time evolution of $i(t)$ and $v(t)$ is reported in Fig.~\ref{fig:timeevolutioniandv}. Starting from the acquired data reported in Figs.~\ref{fig:timeevolutioniandv2} and~\ref{fig:timeevolutioniandv} we have estimated the largest Lyapunov exponents: for the data referring to memristor 1 (Fig.~\ref{fig:timeevolutioniandv2}) the largest Lyapunov exponent is $\lambda_{max}=0.7947$, and for the data from memristor 2 (Fig.~\ref{fig:timeevolutioniandv}) $\lambda_{max}=0.9230$. The existence of positive values for the largest Lyapunov exponent indicates that the behavior is chaotic. As the rest of the set-up was linear, this chaos arises from the memristor, and thus we have demonstrated chaotic dynamics from a single memristor device.

\subsection{Numerical results}
\label{sec:resultsB}
\begin{figure}
\centering
\subfigure[]{{\includegraphics[width=8cm]{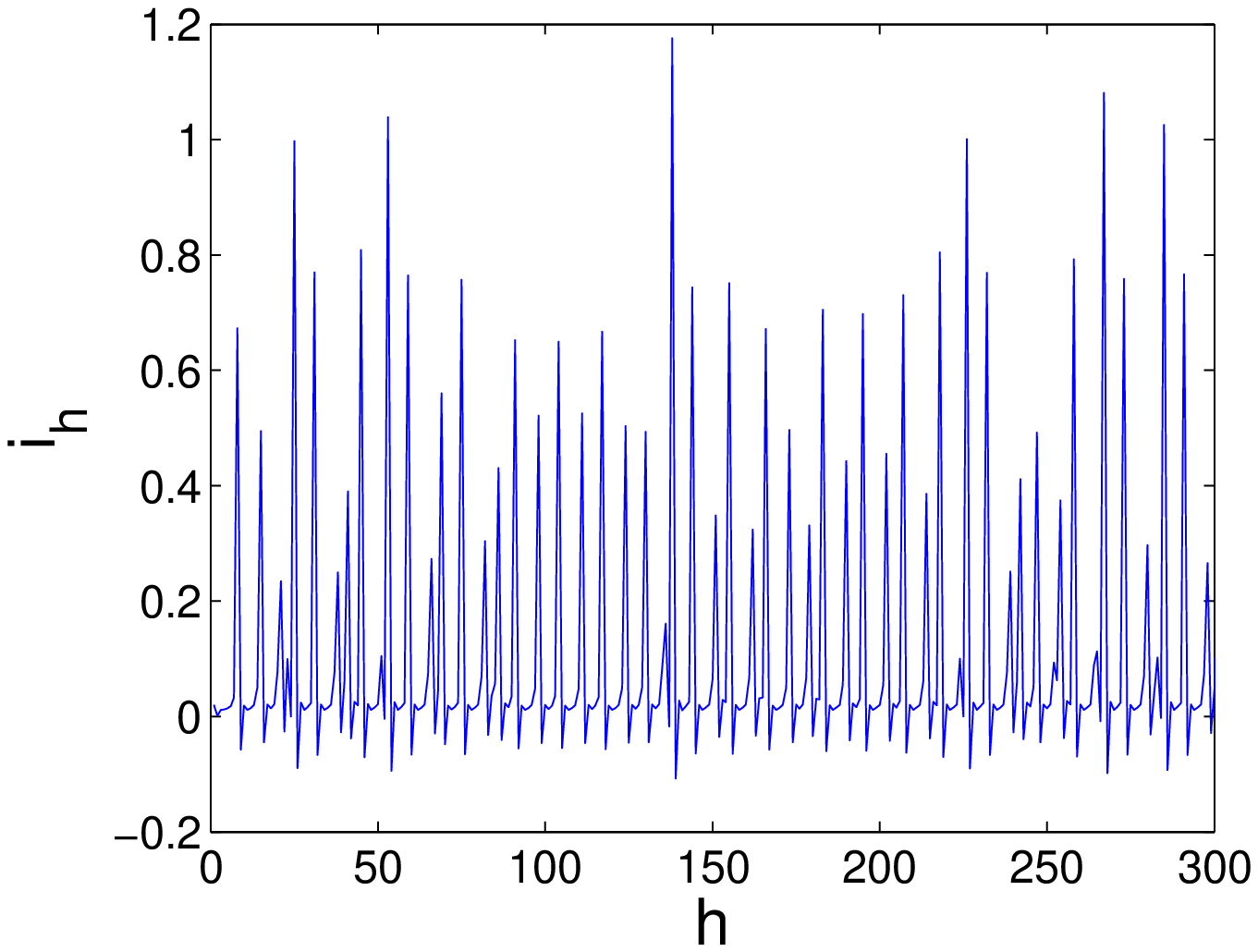}}}
\subfigure[]{{\includegraphics[width=8cm]{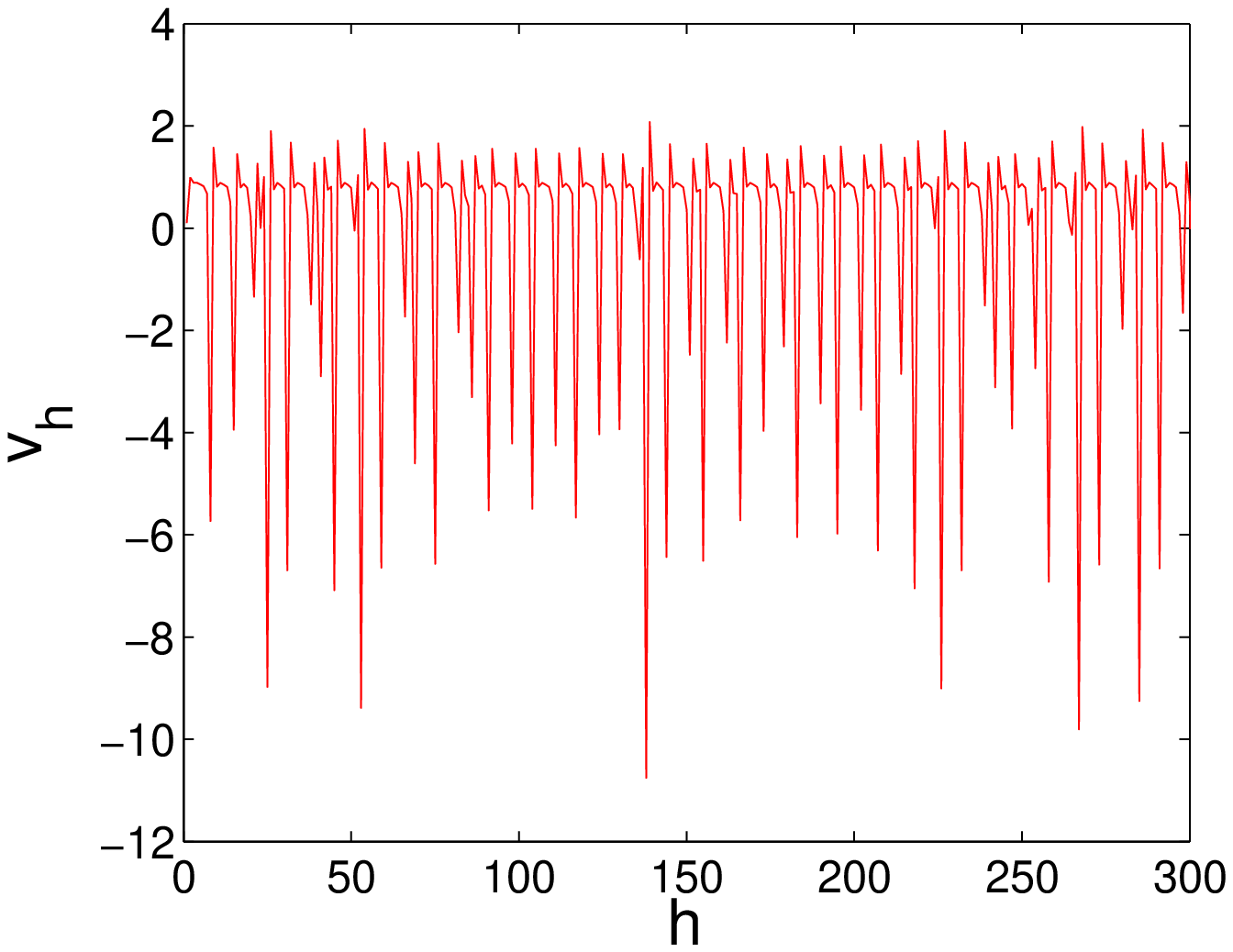}}}
\caption{\label{fig:trend} Trend of the current (a) and voltage (b) of the memristor circuit for $k=10$ and $\beta=100$.}
\end{figure}

\begin{figure}
\centering {{\includegraphics[width=8cm]{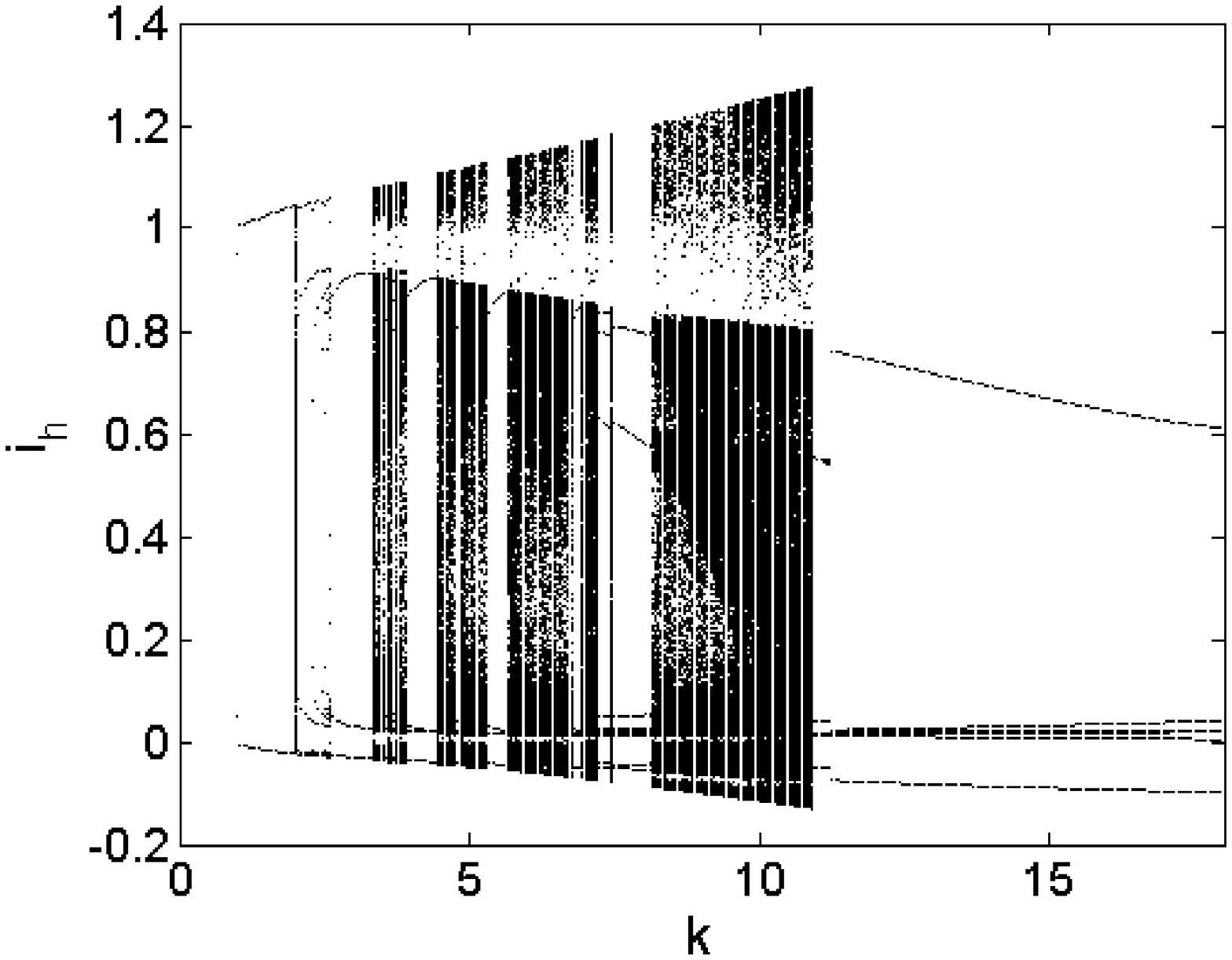}}}
\caption{\label{fig:bifk} Bifurcation diagram of the memristor circuit with respect to $k$ ($\beta=100$, $r=1$).}
\end{figure}

\begin{figure}
\centering {{\includegraphics[width=8cm]{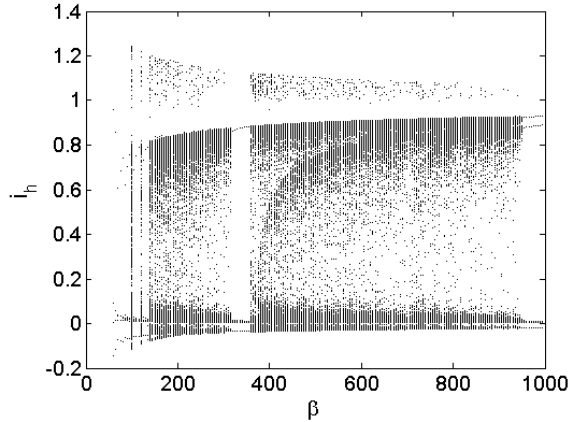}}}
\caption{\label{fig:bifbeta} Bifurcation diagram of the memristor circuit with respect to $\beta$ ($k=10$, $r=1$).}
\end{figure}

In order to demonstrate that the chaos observed with our devices was a real effect generally due to the memristor behaviour, we simulated the set-up and numerically analyzed it with respect to the parameters $k$ and $\beta$. Several regions in the parameter space for which chaotic behavior arises were found. An example of chaotic behavior is reported in Fig.~\ref{fig:trend} where the time evolutions of the variables $v(t)$ and $i(t)$ are shown for $k=10$ and $\beta=100$. In Fig.~\ref{fig:bifk} the bifurcation diagram with respect to $k$ (for $\beta=100$) is illustrated, it shows alternating windows of periodic behavior and chaos. The bifurcation diagram with respect to $\beta$ is reported in Fig.~\ref{fig:bifbeta} (for $k=10$), showing how chaos is preserved for a wide range of values of the constitutive parameter $\beta$. Since $\beta$ may vary in real memristors due to technology, fabrication process and device characteristics, the robustness with respect to this parameter is particularly important.

We observe that the numerical results have been obtained with a dimensionless set of equations; the parameters of the control law ($k$ and $r$) have been rescaled during the experimental tests. The numerical results show that chaos can be generated using this approach. However, the waveforms observed in the experiments and in the numerical simulations do not coincide, indicating that the simplified mathematical model of Eqs. (\ref{eqn:hp model})-(\ref{eq:w2}) is able to capture the general behavior of the system, but not the details of the waveforms observed in the experiments. This shows that the chaotic behaviour comes from an aspect of the memristor that is captured qualitatively by the simplest possible memristor model. However, we strongly suspect that further work involving using more complicated models will produce numerical results closer to experimental data, and this might provide memristor modellers a test case for the improvement of their models.

It is interesting to ask why the memristor allows chaos to arise in such a simple system, and the fact that such a simple model of memristance demonstrates chaotic dynamics provides a clue. As mentioned earlier, the memristor non-linearity is not time-invariant, and it is known that memristors possess a memory. We posit that it is specifically the interaction of this time-varying memory with the time-based feedback that gives rise to chaos in this system (as the voltage for each step is generated on the data from the previous step, there is a memory in the experimental system). The simulation equations are so simple that there is only one time-varying variable, $w(t)$ which is the internal state of the memristor and its memory. This variable depends on $i(t)$ (which is linearly updated through the feedback) but also on the past history of the memristor through a functional which, instead, is non-linear.

\section{Conclusions}
\label{sec:conclusion}
In this work experimental findings on the generation of chaos with a spin-coated memristor are presented. The experimental set-up consists of a single memristor driven by a linear control law relating the applied voltage to the actual value of the current flowing in it. This feedback law does not introduce any non-linearity, so that the observed behavior can be uniquely attributed to the memristor characteristics.
Chaotic waveforms have been experimentally observed in several samples of spin-coated memristors and confirmed by a numerical analysis. The largest Lyapunov exponent has been calculated in the experimental time-series, and the positive value obtained indicates chaotic dynamics. A simple model of memristive behavior has been shown to be sufficient to reproduce the onset of chaos, although capturing all the features of the waveforms observed in the experiments requires a more detailed model. The experiments showing the irregularity of the behavior of the circuit demonstrate the suitability of this simple approach to generate chaos with memristors and suggests that memristors might be useful components for hardware chaos-based encryption.

The analysis performed through the use of the Strukov model suggests that the main mechanism explaining chaos in the experimental and simulation behaviors is the presence of the pinched hysteretic behavior. However, as in the experiments specific SET and RESET processes are not applied, it cannot be excluded the presence of RTN, which on the contrary could be a possible explanation for the differences between experimental and simulation results.

\end{document}